# A New Large-Number Coincidence and a Scaling Law for the Cosmological Constant

Scott Funkhouser
Dept. of Physics, The Citadel, 171 Moultrie St., Charleston SC 29409

ABSTRACT

An ensemble of pure numbers of order near $10^{122}$ is produced naturally from the fundamental parameters of modern cosmology. This new large-number coincidence problem is resolved by demonstrating implicit physical connections that follow from the standard cosmological model. However, the occurrence of the new large-number coincidence combined with the known coincidence among pure numbers of order near $10^{40}$ poses a distinct problem that is resolved with a scaling law for the cosmological constant that was originally proposed by Zel'dovich.

## I. *Introduction*

With the discovery of the accelerating Universe and re-introduction of Einstein's cosmological constant, a new mystery concerning large, pure numbers was initiated. The Planck energy density is roughly 122 orders of magnitude larger than the apparent cosmological vacuum density (Sahni 2000). The discrepancy between the Planck density and the vacuum density is problematic primarily because of its inexplicable enormity. However, the mystery surrounding the factor $10^{122}$ is compounded by the fact that an ensemble of distinct, pure numbers of similar order is produced naturally by the fundamental parameters of modern cosmology.

In Section II of this letter six, physically significant, pure numbers of order near $10^{122}$ are presented. Those pure numbers constitute a compelling, new large-number coincidence problem. The new coincidence problem is then resolved by the analysis in Section III, which demonstrates implicit physical connections among the pure numbers that follow from the standard cosmological model. However, the same basic parameters of nature also generate a coincidence problem that involves pure numbers of order near $10^{40}$. Since it is extremely unlikely for chance alone to cause two separate large-number coincidences to manifest from a few basic parameters, there is good motivation to hypothesize that the pure numbers of order near $10^{122}$ are physically scaled to the pure numbers of order near $10^{40}$. In Section IV it is shown that such a relationship leads to a scaling law between the mass of the nucleon and the cosmological constant that was originally proposed by Zel'dovich.

## II. *The Pure Numbers of Order Near $10^{122}$*

The existence of a new large-number coincidence problem is established here by presenting a collection of pure numbers of order near $10^{122}$. The pure numbers all follow naturally from basic ratios of fundamental parameters and require no arbitrary powers or coefficients. The parameters from which the pure numbers are generated are presented in Table 1, and the resulting large, pure numbers are presented in Table 2. The cosmological parameters result from a relativistic treatment of a flat universe in which the current Hubble parameter $H_0$ is 71km/s/Mpc, the fraction $\Omega_\Lambda$ of the total cosmic energy density due to the vacuum is 0.73 and the fraction $\Omega_m$ due to matter is 0.27. Those parameters are consistent with nominal empirical values (Spergel 2003). Although two significant figures are used for the calculations in this work, the conclusions would all still be valid even if only integer powers of ten were used. If improved observations should eventually indicate different cosmological parameters than the ones employed here, this work would still be valid providing that the estimations of the parameters do not change by more than a few orders of magnitude.

The first two large numbers of order near $10^{122}$ involve the putative cosmological constant, $\Lambda$. The vacuum energy density due to the cosmological constant is $\varepsilon_\Lambda = 3\Lambda c^2/(8\pi G)$, where $c$ is the vacuum-speed of light and $G$ is the Newtonian gravitational coupling. With $\Lambda \approx 1.2\times 10^{-35}\text{s}^{-2}$ the vacuum density is roughly $6.2\times 10^{-10}\text{J/m}^3$. The Planck energy density, defined as $\varepsilon_P \equiv m_P c^2 l_P^{-3}$, where $m_P$ is the Planck mass and $l_P$ is the Planck length, is roughly $4.5\times 10^{113}\text{J/m}^3$. The ratio of the Planck density to the vacuum density is defined here as $n_1$ and is

$$n_1 \equiv \frac{\varepsilon_P}{\varepsilon_\Lambda} \approx 7.1 \times 10^{122}. \tag{1}$$

The pure number in Eq. (1) is proportional to the fundamental pure number $c^5/(G\hbar\Lambda)$, where $\hbar$ is the Planck constant. The Wesson mass $m_W \equiv c^{-2} h\sqrt{\Lambda/3}$, where $h \equiv 2\pi\hbar$, is roughly $1.5\times 10^{-68}\text{kg}$ and it is the smallest, physically significant quantum of mass that could be associated with our cosmos (Wesson 2004). The largest possible mass that could be associated with our current universe is the observable mass, which is the mass $M_{p0}$ contained within the sphere whose radius is the current particle horizon $R_{p0}$. The mass of the observable universe is roughly $9.3\times 10^{53}\text{kg}$. The ratio $n_2$ of the largest mass to the smallest mass is thus

$$n_2 \equiv \frac{M_{p0}}{m_W} \approx 6.3 \times 10^{121}. \tag{2}$$

The next three large numbers presented are related to the cosmic capacities for information and computation. According to the Bekenstein-Hawking entropy bound the maximum number of degrees of freedom available to the Universe is one fourth of the surface area $4\pi R_e^2$ of the sphere whose radius is the cosmic event horizon $R_e$, measured in Planck units (Bousso 2002). In this current age the maximum number of degrees of freedom, defined here as $n_3$, is

$$n_3 \equiv \frac{\pi R_{e0}^2}{l_P^2} \approx 2.5 \times 10^{122}, \tag{3}$$

where $R_{e0}$ is the current event horizon. According to the Margolus/Levitin theorem the maximum rate at which logical operations could be performed by a physical system with energy $E$ is $2E/(\pi\hbar)$ (Margolus 1998). Thus, the maximum number of operations that could have been performed by the mass of the observable Universe is

$$n_4 \equiv \frac{2M_{p0}c^2 T_0}{\pi\hbar} \approx 2.1 \times 10^{122}, \tag{4}$$

where $T_0$ is the age of the Universe (Lloyd 2002). The fifth pure number of order near $10^{122}$ presented here follows from holographic considerations applied to the nucleon (Mena 2002). The characteristic size of the nucleon is of order near $10^{-15}$m, which is of order of the Compton wavelength of the nucleon $l_n \equiv h/(m_n c)$. The number of nucleon volumes $V_n \sim l_n^3$ contained within the volume $V_{e0} \sim R_{e0}^3$ of the sphere whose radius is the current event horizon is

$$n_5 \equiv \frac{V_{e0}}{V_n} \approx 1.3 \times 10^{123}. \tag{5}$$

Another pure number contributing to the new coincidence problem is found by considering the fundamental scales of gravitational energy. The gravitational potential energy of the cosmic mass is roughly $GM_{p0}^2/R_{p0}$. The scale of gravitational binding energy associated with a nucleon is roughly $Gm_n^2/l_n$. The ratio of the two, defined as $n_6$, is

$$n_6 \equiv \frac{M_{p0}^2/R_{p0}}{m_n^2/l_n} \approx 9.3 \times 10^{119} \tag{6}$$

Finally, there are at least two other pure numbers of order near $10^{122}$ that may be produced naturally from the parameters of the universe:

$$\frac{GM_{e0}^2}{\hbar c} \approx 2.4 \times 10^{120} \tag{7}$$

and

$$\frac{M_{e0}R_{e0}c}{\hbar} \approx 1.4 \times 10^{121}. \tag{8}$$

The terms in (7) and (8) do not have any known physical significance and therefore should not be considered to contribute strongly to the new coincidence problem.

The pure numbers $n_1$ through $n_6$ constitute a new large-number coincidence. It is unlikely for chance alone to be responsible for generating so many similar pure numbers from just several fundamental parameters. It is reasonable to hypothesize that the new coincidence problem among large, pure numbers of order near $10^{122}$ results from some underlying physics rather than chance alone. That hypothesis is verified in the following section.

III. *Resolving the new coincidence problem*

The coincidence problem associated with the pure numbers of order near $10^{122}$ is the result of a few physical scaling laws that follow from the standard model of cosmology. The Friedmann-Robertson-Lemaitre-Walker (FRLW) equations summarize the general relativistic component of the model. They provide a relativistic treatment of the expansion of the universe by relating the dynamics of the cosmic scale factor to the average density of energy in the cosmos. In a flat universe the primary FRLW equation is

$$\left(\frac{\dot{a}}{a}\right)^2 = \frac{8\pi G}{3c^2}(\varepsilon_m + \varepsilon_\Lambda + \varepsilon_r), \tag{9}$$

where $a$ is the scale factor, $\dot{a}$ is its rate of change, $\varepsilon_m$ is the average density of mass-energy and $\varepsilon_r$ is the density of energy due to photons. The density of photons is negligible now since the era of radiation-dominance ended very long ago. The density $\varepsilon_m$ of matter varies with the scale factor according to $\varepsilon_m = \varepsilon_{m0}a^{-3}$, where $\varepsilon_{m0}$ is the current density of matter (and the current scale factor $a_0$ is defined to be one). The vacuum density $\varepsilon_\Lambda$ is treated as if it were constant. The left side of Eq. (9) is defined to be $H^2$, where $H$ is the Hubble parameter. (Bergstrom 2006)

By virtue of the cosmic coincidence the densities $\varepsilon_\Lambda$ and $\varepsilon_{m0}$ are now of the same order. It follows immediately from the cosmic coincidence and Eq. (9) that the current

Hubble parameter $H_0$ is near $\Lambda^{1/2}$. The age $T$ of the universe associated with some scale factor $a$ can be obtained from Eq. (9) by the integration

$$T = \int_0^a \frac{d\alpha}{\dot{\alpha}}, \qquad (10)$$

where $\alpha$ is the variable of integration representing the scale factor. During matter-dominance, and still roughly at this time, the age of the universe $T$ and the inverse of the Hubble parameter, $H^{-1}$, are proportional to $a^{3/2}$, and thus are mutually scaled. Since $H_0 \sim \Lambda^{1/2}$, the current age $T_0$ of the universe is scaled to the cosmological constant:

$$T_0 \sim \Lambda^{-1/2}. \qquad (11)$$

The particle horizon $R_p$ associated with some scale factor $a$ may be obtained from Eq. (9) by the integration

$$R_p = a \int_0^a c \frac{d\alpha}{\dot{\alpha}\alpha}. \qquad (12)$$

During matter-dominance the particle horizon is proportional to $a^{3/2}$, and thus scaled to the age of the universe $T$. Therefore, due to the cosmic coincidence and Eq. (11), the current particle horizon is scaled to $c\Lambda^{-1/2}$. The event horizon $R_e$ associated with some scale factor $a$ is given by the integration

$$R_e = a \int_a^\infty c \frac{d\alpha}{\dot{\alpha}\alpha}. \qquad (13)$$

In times just before vacuum-dominance and at all times after, the event horizon is of the order the De Sitter horizon $R_\Lambda \equiv c(3/\Lambda)^{1/2}$, which is the maximum event horizon that may exist in a universe with a positive cosmological constant. Thus, due to the cosmic coincidence, the current event horizon $R_{e0}$ is of the order the particle horizon $R_{p0}$, and both are of the order the De Sitter horizon

$$R_{p0} \sim R_{e0} \sim c\Lambda^{-1/2}. \qquad (14)$$

The event horizon is asymptotically approaching the De Sitter horizon. The particle horizon, however, will increase without limit, becoming proportional to the scale factor.

The observable mass $M_p$ is defined as the average density of matter in the universe multiplied by the volume of the sphere whose radius is the particle horizon. During matter-dominance the volume of the particle sphere is proportional to $a^{9/2}$. Since the density of matter is proportional to $a^{-3}$ the observable mass $M_p$ is proportional to $a^{9/2} a^{-3} = a^{3/2}$, and thus proportional to the particle horizon $R_p$, the age $T$ and the inverse of the Hubble parameter, $H^{-1}$. Specifically, during the era of matter-dominance and still roughly at this time, the following scaling law is satisfied:

$$M_p \sim \frac{c^2}{G} R_p. \qquad (15)$$

Due to the cosmic coincidence and Eq. (14), the current observable mass $M_{p0}$ is scaled to the current mass $M_{e0}$ of the sphere whose radius is the event horizon, and both are scaled to the cosmological constant according to

$$M_{p0} \sim M_{e0} \sim \frac{c^3}{G\Lambda^{1/2}}. \qquad (16)$$

Eqs. (11)-(16) constitute the underlying physics responsible for generating the coincidence problem associated with the pure numbers in Eqs. (1) – (8). The number $n_2$ is scaled to $n_1$ due to Eq. (16). The number $n_3$ is scaled to $n_1$ due to Eq. (14). (As a result of the cosmic coincidence the maximum number of degrees of freedom in the Universe is currently of the order the maximum number that the Universe could ever contain, which is roughly $(R_\Lambda/l_P)^2 \sim c^5/(G\hbar\Lambda)$, which is proportional to $n_1$). The number $n_4$ is scaled to $n_1$ due to Eqs. (11) and (16). The number $n_5$ reduces to $n_6$ due to Eqs. (14) and (16). Though they were not formally considered to be part of the new large-number coincidence, the terms in Eqs. (7) and (8) are scaled to $n_1$ due to Eqs. (14) and (16). All that remains of the coincidence problem among pure numbers of order near $10^{122}$ is the similarity between $n_1$ and $n_6$, which no longer constitutes a compelling problem *per se*. The new large-number coincidence is therefore the result of the physics of the standard cosmological model and the cosmic coincidence.

IV. *Two Large-Number Coincidences and the Cosmological Constant*

The new coincidence problem among pure numbers of order near $10^{122}$ is resolved. However, its existence is still problematic given that there is another large-number coincidence that is generated from the fundamental parameters of the universe. That other large-number coincidence originates with Weyl, Dirac and Eddington, and it concerns pure numbers of order near $10^{40}$ (Dirac 1974). The existence of that older coincidence problem motivated the formulation of the so-called Large Numbers Hypothesis, which suggests that the large, pure numbers that are formed by the parameters of nature must be related physically (Dirac 1974). The prevalence of two large, pure numbers among the parameters of nature poses a special problem that begs explanation. Whether or not the two large-number coincidences occur only in this epoch is not important since their coexistence would be a problem even if they occurred together in every epoch.

For the purposes of this present investigation the known, physically significant pure numbers of order near $10^{40}$ are presented below and listed in Table 3. The gallery of such terms presented here differs slightly from the collection usually presented in this context (Funkhouser, 2006). Some of the pure numbers of order near $10^{40}$ that have been addressed historically involved arbitrary powers, and some terms bore no direct physical significance. Such terms have been excluded from the ensemble presented here. Two new pure numbers of order near $10^{40}$ are presented here as well.

The known, physically significant pure numbers of order near $10^{40}$ are as follows. The maximum number of degrees of freedom allowed within a sphere whose radius is the Compton wavelength of the nucleon is defined to be $j_1$ and is

$$j_1 \equiv \pi \frac{l_n^2}{l_P^2} \approx 2.1 \times 10^{40}. \tag{17}$$

The term $j_1$ in Eq. (17) is also roughly the ratio of the rest energy of the nucleon to its characteristic gravitational energy $Gm_n^2/l_n$. The gravitational potential energy of a nucleon due to the mass distribution of the cosmos is roughly $GM_{p0}m_n/R_{p0}$. The characteristic gravitational binding energy of the nucleon is roughly $Gm_n^2/l_n$. The ratio of the two is defined to be $j_2$ and is

$$j_2 \equiv \frac{M_{p0}/R_{p0}}{m_n/l_n} \approx 1.7 \times 10^{39} \tag{20}$$

The ratio of the electrostatic force between the electron and proton to the gravitational force between them is

$$j_3 \equiv \frac{k_e e^2}{G m_n m_e} \approx 2.3 \times 10^{39}, \tag{20}$$

where $m_n$ represents the proton mass, $m_e$ is the electron mass, $k_e$ is the Coulomb constant and $e$ is the fundamental charge unit. The maximum number of logical operations that could have been performed by a single nucleon over the age of the universe is, according to the Margolus/Levitin theorem,

$$j_4 \equiv \frac{2 m_n c^2 T_0}{\pi \hbar} \approx 3.9 \times 10^{41}. \tag{18}$$

The term $j_4$ is proportional to ratio of the age of the universe to the time required for light to traverse the Compton wavelength of the nucleon. The ratio of the current particle horizon to the Compton wavelength of the nucleon is

$$j_5 \equiv \frac{R_{p0}}{l_n} \approx 3.3 \times 10^{41}. \tag{19}$$

Finally, the ratio of the nucleon mass to the smallest possible mass, being the Wesson mass, is

$$j_6 \equiv \frac{m_n}{m_W} \approx 1.1 \times 10^{41}. \tag{21}$$

The terms $j_1$ through $j_6$ constitute conservatively the coincidence problem among pure numbers of order near $10^{40}$. The number $j_5$ is scaled to $j_6$ because of Eq. (14). The number $j_4$ is scaled to $j_6$ due to Eq. (11). The number $j_3$ is similar to $j_1$ since $m_e/m_n$ is not very different than the fine structure constant. The number $j_2$ reduces to $j_6$ due to Eqs. (14) and (16). All that remains is the similarity between $j_1$ and $j_6$, which does not constitute a compelling coincidence problem. The Dirac-Eddington large-number coincidence is thus resolved. Note that this resolution of the problem alleviates the need to introduce time-variation of parameters such as $G$. (Funkhouser 2006)

Even though both the coincidence problem among pure numbers of order near $10^{40}$ and the coincidence problem among pure numbers of order near $10^{122}$ have been resolved, it is yet problematic that two such coincidence problems are generated from the same set of parameters. There are also several remarkable algebraic similarities between the two sets of large, pure numbers that compound that coincidence. First, note that $j_4$, $j_5$ and $j_6$ are all scaled to $n_6^{1/3}$ due to Eqs. (11), (14) and (16). That does not mean that the two coincidence problems are identical, since $n_6$ does not scale to $n_1$, $n_2$, $n_3$ or $n_4$, which are all mutually scaled. However, the unresolved similarity between $j_1$ and $j_6$ among the pure numbers of order near $10^{40}$ is algebraically equivalent to the unresolved similarity between $n_1$ and $n_6$ among the pure numbers of order near $10^{122}$. (The numbers $j_1$, $j_2$ and $j_3$ do not scale to any of the pure numbers of order near $10^{122}$ through the standard cosmological model.) Furthermore, the two coincidence problems are resolved through the same basic physics from the standard cosmological model, being Eqs. (9) – (16). Most suggestive is the fact that all of the known pure numbers of order near $10^{40}$ would be scaled to all of the known pure numbers of order near $10^{122}$ if $j_1$ were scaled to $j_6$, or, equivalently, if $n_1$ were scaled to $n_6$.

In order to resolve the host of coincidences associated with the large pure numbers it is well motivated to propose that the pure numbers of order near $10^{122}$ are proportional to the third power of the pure numbers of order near $10^{40}$. Conveniently, that is equivalent to hypothesizing that $j_1 \sim j_6$, or, rather, $n_1 \sim n_6$. Such a relationship would cause the two large-number coincidences to be identical and would amount to the scaling law

$$m_n \sim \left(\frac{h^4 \Lambda}{G^2 c^2}\right)^{1/6}. \tag{22}$$

The term on the right side of Eq. (22) is roughly $4.2 \times 10^{-28}$kg, which is within an order of magnitude of the nucleon mass scale. The relation in Eq. (22) was proposed to explain why the coincidence among pure numbers of order near $10^{40}$ occurs simultaneously with the cosmic coincidence (Funkhouser 2006). Eq. (23) is essentially the same scaling law first proposed by Zel'dovich based on considerations of field theory (Zel'dovich 1967). Mena Marugan and Carneiro have proposed that same relationship based on holographic considerations (Mena 2002). The scaling law in Eq. (22) also follows naturally from applying the Bekenstein-Hawking bound to a quantum-cosmological model of a three-dimensional universe inflating from the collapse of seven extra dimensions (Funkhouser 2007).

Lloyd has already shown that the maximum number $N_\Lambda$ of allowed bits in our cosmos is related to certain products of the pure numbers of order near $10^{40}$ regardless of any underlying relationship (Lloyd 2002). If the scaling law in Eq. (22) should be physical then $N_\Lambda$ would be scaled to the product of any three of the known pure numbers of order near $10^{40}$. That is because all of the known pure numbers of order near $10^{40}$ would be mutually scaled if Eq. (22) should be physically meaningful.

Table 1; List of cosmological parameters discussed in this work

| Parameter | Approximate Value | Description |
|---|---|---|
| $\Lambda$ | $1.2 \times 10^{-35}$s$^{-2}$ | cosmological constant |
| $\varepsilon_\Lambda$ | $6.2 \times 10^{-10}$J/m$^3$ | vacuum density |
| $R_{e0}$ | $1.5 \times 10^{26}$m | current event horizon |
| $R_{p0}$ | $4.4 \times 10^{26}$m | current particle horizon |
| $M_{e0}$ | $3.4 \times 10^{52}$kg | current mass of event sphere |
| $M_{p0}$ | $9.3 \times 10^{53}$kg | current mass of particle sphere |
| $T_0$ | $4.3 \times 10^{17}$s | current age of the universe |

Table 2; Pure numbers of order near $10^{122}$

| Name | Definition | Approximate Value |
|---|---|---|
| $n_1$ | $\varepsilon_P / \varepsilon_\Lambda$ | $1.7 \times 10^{122}$ |
| $n_2$ | $M_{p0} / m_W$ | $6.3 \times 10^{121}$ |
| $n_3$ | $\pi R_{e0}^2 / l_P^2$ | $2.5 \times 10^{122}$ |
| $n_4$ | $2 M_{p0} c^2 T_0 / (\pi \hbar)$ | $2.1 \times 10^{122}$ |
| $n_5$ | $V_{e0} / V_n$ | $1.3 \times 10^{123}$ |
| $n_6$ | $M_{p0}^2 l_n / (m_n^2 R_{p0})$ | $9.3 \times 10^{119}$ |

Table 3; Pure numbers of order near $10^{40}$

| Name | Definition | Approximate Value |
|---|---|---|
| $j_1$ | $\pi l_n^2 / l_P^2$ | $2.1 \times 10^{40}$ |
| $j_2$ | $M_{p0} l_n / (m_n R_{p0})$ | $1.7 \times 10^{39}$ |
| $j_3$ | $k_e e^2 / (G m_n m_e)$ | $2.3 \times 10^{39}$ |
| $j_4$ | $2 m_n c^2 T_0 / (\pi \hbar)$ | $3.9 \times 10^{41}$ |
| $j_5$ | $R_{p0} / l_n$ | $3.3 \times 10^{41}$ |
| $j_6$ | $m_n / m_W$ | $1.1 \times 10^{41}$ |